# Vapor-Solid Growth of High Optical Quality MoS$_2$ Monolayers With Near-Unity Valley Polarization

Sanfeng Wu[1*], Chunming Huang[1*], Grant Aivazian[1], Jason S Ross[2], David H Cobden[1], Xiaodong Xu[1,2†]

[1] Department of Physics, University of Washington, Seattle, Washington 98195, USA

[2] Department of Material Science and Engineering, University of Washington, Seattle, Washington 98195, USA

* These authors contribute equally to this work.

† Email address: xuxd@uw.edu

**Abstract**

**Monolayers of transition metal dichalcogenides (TMDCs) are atomically thin direct-gap semiconductors with potential applications in nanoelectronics, optoelectronics, and electrochemical sensing. Recent theoretical and experimental efforts suggest that they are ideal systems for exploiting the valley degrees of freedom of Bloch electrons. For example, Dirac valley polarization has been demonstrated in mechanically exfoliated monolayer MoS$_2$ samples by polarization-resolved photoluminescence, although polarization has rarely been seen at room temperature. Here we report a new method for synthesizing high optical quality monolayer MoS$_2$ single crystals up to 25 microns in size on a variety of standard insulating substrates (SiO$_2$, sapphire and glass) using a catalyst-free vapor-solid growth mechanism. The technique is simple and reliable, and the optical quality of the crystals is extremely high, as demonstrated by the fact that the valley polarization approaches unity at 30 K and persists at 35% even at room temperature, suggesting a virtual absence of defects. This will allow greatly improved optoelectronic TMDC monolayer devices to be fabricated and studied routinely.**

The transition metal dichalcogenides MX$_2$ (M=Mo,W; X=S, Se, *etc*) have layered structures with van der Waals interactions between the layers. Monolayers of such materials were first obtained by the mechanical exfoliation technique typically used for graphene.[1] Subsequent investigation has shown that these two-dimensional (2D) semiconductors[1,2,3] exhibit unique properties, such as a transition from an indirect bandgap in the bulk to a direct bandgap at monolayer thicknesses, [3, 4] massive Dirac-like behavior of the electrons, [5] excellent field-effect transistor performance at room temperature, [6] and completely tunable 2D excitonic effects. [7]

Recently, these monolayers have also been suggested as good candidates for the realization of valley-based electronics. [5, 8, 9, 10] In monolayer $MoS_2$ there are two energy-degenerate Dirac valleys at the corners of the hexagonal Brillion zone. [5, 10] The Berry curvature and magnetic moments of electrons associated with different valleys have opposite sign and are linked to measurable quantities which can distinguish the valleys, such as k-resolved optical dichroism, offering the possibility of manipulating and utilizing the valley degree of freedom. [11, 12] Valley polarization has been demonstrated in $MoS_2$ monolayers by circularly polarized light excitation, [8, 9, 10] and electrical control of it has been reported in bilayer samples. [13]

Progress thus far has relied mainly on mechanically exfoliated samples where scaling for device applications [14] is probably impossible. Recent attempts to develop more scalable techniques include exfoliation in liquids, [2, 15, 16] hydrothermal synthesis, [17] epitaxy growth using graphene, [18] and soft sulfurization. [19, 20] However, these methods are not easily integrated with device fabrication. Chemical vapor deposition has also been explored using an Mo film [21] (or $MoO_3$ powder [22]) and sulfur powder as the reactants, yielding monolayers of $MoS_2$ on 300nm $SiO_2$/Si substrate compatible with device fabrication. [21, 22] It has yet to be proven though that such monolayers have sufficient quality for investigating valley-related physics. Inter-valley scattering enhanced by defects and impurities can reduce or destroy the valley polarization, as evident from the disparate degrees of polarization reported by different groups. [8, 9, 10, 13, 23] A high degree of valley polarization is required for valley physics and is also a hallmark of crystal quality.

Here we introduce a new and straightforward method for obtaining high optical quality monolayer $MoS_2$ *via* a vapor-solid (VS) growth mechanism. [24] Up to 400 $\mu m^2$ monolayer flakes with triangular shape are directly produced on insulating substrates such as $SiO_2$, sapphire, and glass, without using any catalysts. The growth procedure is simple physical vapor transport, using an $MoS_2$ powder source and Ar carrier gas (details are given in Fig. 1 and Methods), similar to the procedure used for growing $Bi_2Se_3$ topological insulator nano-plates. [24] Using polarization-resolved photoluminescence (PL), [13] we observe valley polarization approaching nearly unity at low temperature (30 K) and 35% at room temperature. This observation demonstrates that these monolayers are of high quality and are suitable for valley physics and applications.

**Results and Discussion**

The resulting $MoS_2$ monolayers are characterized by optical microscopy (OM, Zeiss Axio Imager A1), atomic force microscopy (AFM, Veeco Dimension 3100), scanning electron microscopy (SEM, FEI Sirion), and micro-Raman spectroscopy (Renishaw inVia Raman Microscope). Figure 2 is a typical SEM image of a sample grown on $SiO_2$/Si. The crystallites have lateral dimensions up to 25 microns, and are approximately equilateral triangles (see Fig. 2 inset). This is consistent with the triangular symmetry of monolayer $MoS_2$ (Fig.1c). It suggests that each is a single crystal without extended defects or grain boundaries;[25, 26] the facets are then the most slow-growing or stable symmetry-equivalent crystal planes – it remains to be established whether these are the "zigzag" or the "armchair" edges. Therefore another advantage over exfoliation techniques is that the crystal axes can be immediately identified by inspection.

**Optical and AFM characterization.** Figures 3a-c show optical microscope images of growths on sapphire, glass, and 300 nm $SiO_2$/Si substrates, respectively. The color contrast of all the larger crystallites is uniform; moreover, for those on $SiO_2$/Si (Fig. 3c) it is identical to that of exfoliated monolayers on the same substrate. These facts strongly indicate that they are monolayers.[3, 27] The growth on sapphire is much denser than that on both $SiO_2$ and glass, but on all the substrates nucleation appears to be random, as was found for VS growth of topological insulators.[24] Smaller (<2 μm), thicker crystallites are also present, especially on $SiO_2$. We speculate that the growth kinetics are such that a monolayer is favored, and grows rapidly, if the nucleating crystal is aligned suitably with the substrate; otherwise more three-dimensional growth occurs. The monolayer thickness is confirmed by atomic force microscopy (AFM).[3, 6, 28] Figure 3d shows an AFM image of one crystallite on $SiO_2$, revealing a flat, uniform surface. A line cut along the red line (Fig 3e) shows an apparent thickness of ~0.75 nm on $SiO_2$/Si substrate, consistent with previous measurements of monolayers. [4, 6] Similar measurements on sapphire substrates are shown in the Supplementary Materials.

**Raman characterization.** The samples were also studied by Raman spectroscopy. Typical Raman spectra from the triangular crystallites grown on different substrates, as well as from exfoliated monolayer and bulk MoS$_2$, using a 514.5 nm excitation laser, are shown in Fig. 4. We observe both of the Raman modes ($E_{2g}^1$, and $A_{1g}$) expected for monolayer MoS$_2$. [3, 4, 28, 29] The $E_{2g}^1$ peak is at 386 cm$^{-1}$ for the SiO$_2$ substrate, 384 cm$^{-1}$ for sapphire and 385 cm$^{-1}$ for glass. The $A_{1g}$ peak is at 405 cm$^{-1}$ for SiO$_2$ and sapphire and 404 cm$^{-1}$ for glass. The peak separations are 19 cm$^{-1}$, 21 cm$^{-1}$, and 19 cm$^{-1}$, respectively. All these numbers agree well with the exfoliated monolayer sample. In order to investigate the uniformity of the grown monolayer sample we also performed scanning Raman measurements (excited by a 532nm laser line). Intensity and peak-position maps for a triangular crystallite are show in the Fig. 4b, c, d and e. Note that the different excitation energy leads to a different intensity ratio between the two peaks. The peak separation resulting from the map is $22 \pm 1.5$ cm$^{-1}$. It clearly demonstrates that this entire crystallite is a uniform monolayer.

**Optical valley-selective effect.** To investigate the potential of these monolayer crystallites for 2D optoelectronics and valley-related device applications we performed polarization-resolved PL. [9, 8, 13] Circularly polarized PL measurements can identify valley polarization in monolayer MoS$_2$ created by appropriate optical pumping: the $+K$ and $-K$ valleys are selectively excited by $\sigma^+$ or $\sigma^-$ light respectively, as indicated Fig 5a. [5, 10] Due to the large k-space separation of the valleys, inter-valley scattering is suppressed and the valley relaxation time is longer than the electron-hole recombination time. Emission from a given valley is also circularly polarized and the degree to which the PL has the same helicity as the incident light therefore reflects the degree of valley polarization. Large valley polarization provides evidence for good sample quality, as impurities and defects in the crystal will cause intervalley scattering even at low temperature. [9]

In our measurements, a 632 nm He-Ne laser beam is circularly polarized by a quarter-wave plate (QWP) and focused at normal incidence onto the monolayer sample held in a cryostat. The PL signal is selectively detected for both $\sigma^+$ and $\sigma^-$ polarization using the setup described in Ref. 13.

The laser spot size is about 2 μm with an intensity of ~150 W/cm$^2$. We define the degree of PL polarization, which reflects the valley polarization, as [8, 9, 13] $\eta = \frac{\text{PL}(\sigma^+) - \text{PL}(\sigma^-)}{\text{PL}(\sigma^+) + \text{PL}(\sigma^-)}$.

For a substrate temperature of 30 K, the PL spectra for $\sigma^+$ excitation are shown in Figs. 5b and d for monolayer crystallites on SiO$_2$ and sapphire substrates. The results with $\sigma^-$ excitation are similar (see Supplementary Materials). The cutoff in the spectra at ~ 1.92eV is due to the notch filter placed in the collection optical path for blocking the laser light. The sharp spikes superimposed on the spectra are Raman scattering peaks. The spectra show only a single emission peak at ~1.9eV in SiO$_2$ substrate, in contrast with reports on exfoliated samples [30] where a second broad impurity peak is present at ~1.77eV. The absence of an impurity peak is powerful evidence of excellent crystal quality. [30]

The PL signal is highly $\sigma^+$-polarized for both substrates. Reported degrees of valley polarization at low temperatures from mechanically exfoliated monolayers in the literature vary widely: 30%, [9] 50%, [10] 80%, [13] and up to 100% on boron nitride substrate, [8] showing that intervalley scattering is very sensitive to sample details. The degree of polarization in our monolayers is plotted in Figs. 5c and 5e for both SiO$_2$ and sapphire substrates. We see nearly unity polarization on SiO$_2$ and more than 95% on sapphire, with the polarization decreasing at lower photon energies as in previous reports. [8, 9]

Interestingly, the PL polarization is substantial even at room temperature, approaching a maximum of 35% at ~1.92 eV on both substrates, as shown in Fig. 6. Inter-valley scattering increases with temperature due to enhanced phonon populations, [5] resulting in the decrease of the valley polarization and usually making it vanish at room temperature, [9] although recently [23] there has been a report 40% of valley polarization at 300 K from a mechanical exfoliation sample. Thus our VS grown samples are as good as the highest optical quality samples obtained by mechanical exfoliation.

**Conclusion**

In summary, we report a simple method for growing high optical quality monolayer MoS$_2$ directly on various insulating substrates, which should facilitate device fabrication without the need for a transfer process. The absence of impurity luminescence and the substantial room temperature polarization imply excellent crystal quality and the potential for optoelectronic applications without the need for low temperatures. The technique could also be applicable to other TMDCs.

## Methods

An MoS$_2$ powder source (Alfa Aesar, Purity 99%) in an alumina boat is placed in the center of a horizontal quartz tube furnace (CARBOLITE 12/600 1200C Tube Furnace with 1 inch tube diameter), as illustrated in Fig. 1. The insulating substrate (either 300 nm SiO$_2$/Si, (0001) sapphire, or normal glass) is cleaned in acetone, isopropyl alcohol, and deionized water and is placed downstream far from the oven center in a cooler zone (at ~ 650 $^o$C during growth). The tube is initially pumped to a base pressure of 20 mTorr and flushed with the Ar carrier gas (~20 sccm) repeatedly at room temperature to remove oxygen contamination. With the carrier gas flowing and the pressure maintained at ~20 Torr, the furnace temperature is then increased to ~ 900 $^o$C ( ~ 35 $^o$C/min) and held there for 15 - 20 minutes before being allowed to cool naturally (see supplementary materials).

## Acknowledgments

This work was mainly supported by the U.S. Department of Energy, Office of Basic Energy Sciences, Division of Materials Sciences and Engineering (Awards DE-SC0008145 and DE-SC0002197). GA was supported by DARPA N66001-11-1-4124.

## Supporting Information

The temperature profile of the growth, AFM characterization on the sapphire substrate, and complementary data for the PL polarization are shown in supplementary material. This material is available free of charge *via* the Internet at http://pubs.acs.org.

**Conflict of interest**

The authors declare no competing financial interests

102103.

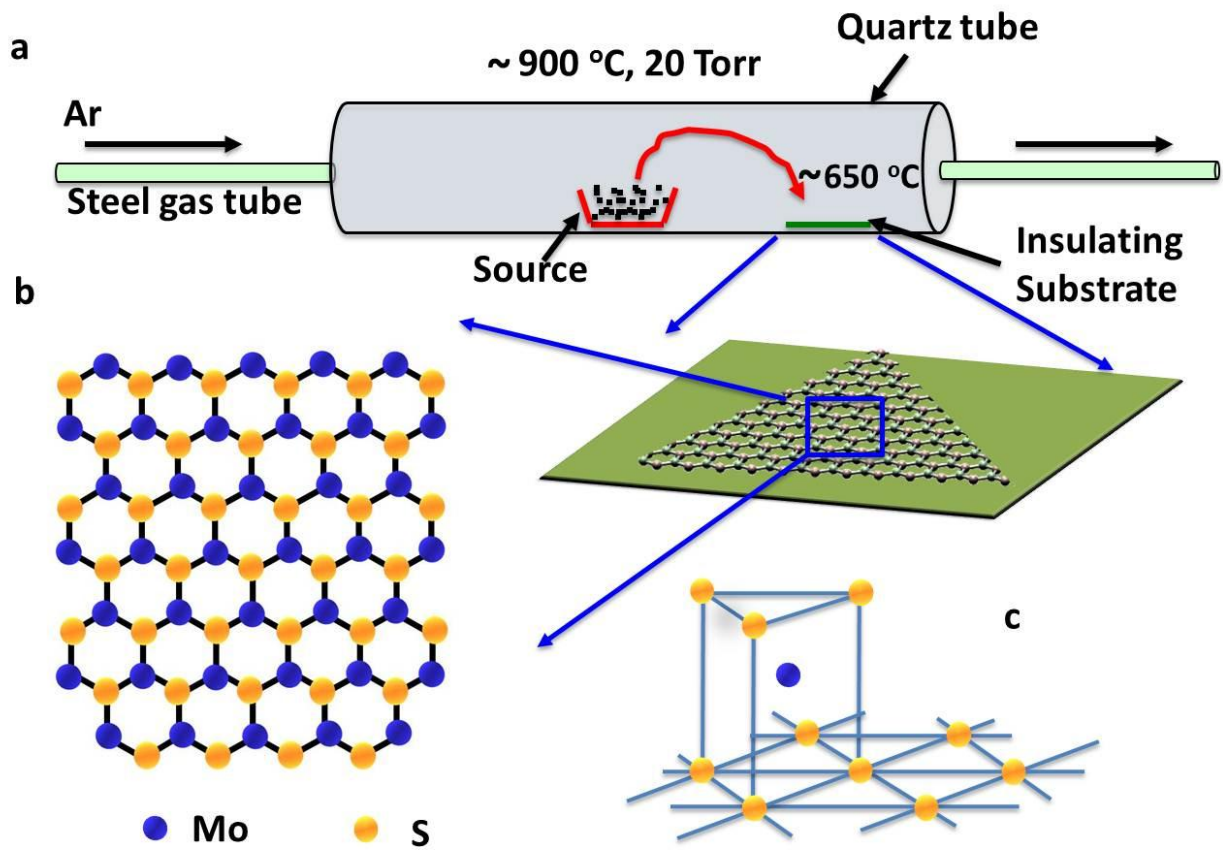

**Figure 1.** (a) Growth setup and conditions. (b) Cartoon indicating the structure of the triangular monolayer crystallites. (c) Structure of monolayer $MoS_2$.

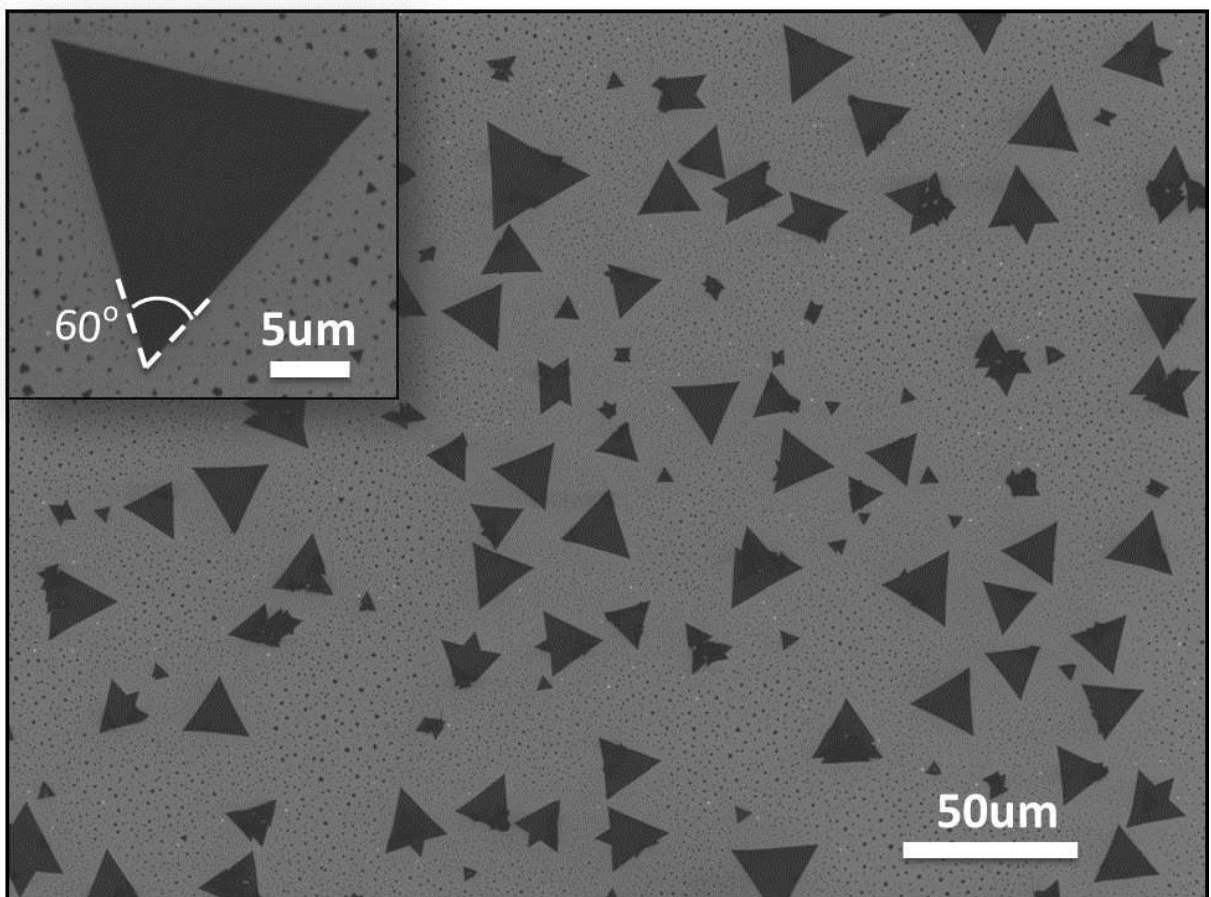

Figure 2. Scanning electron microscope image of triangular MoS$_2$ monolayer crystallites grown on a 300 nm SiO$_2$/Si substrate. The inset shows the 60° corners of a selected crystallite with a clean surface.

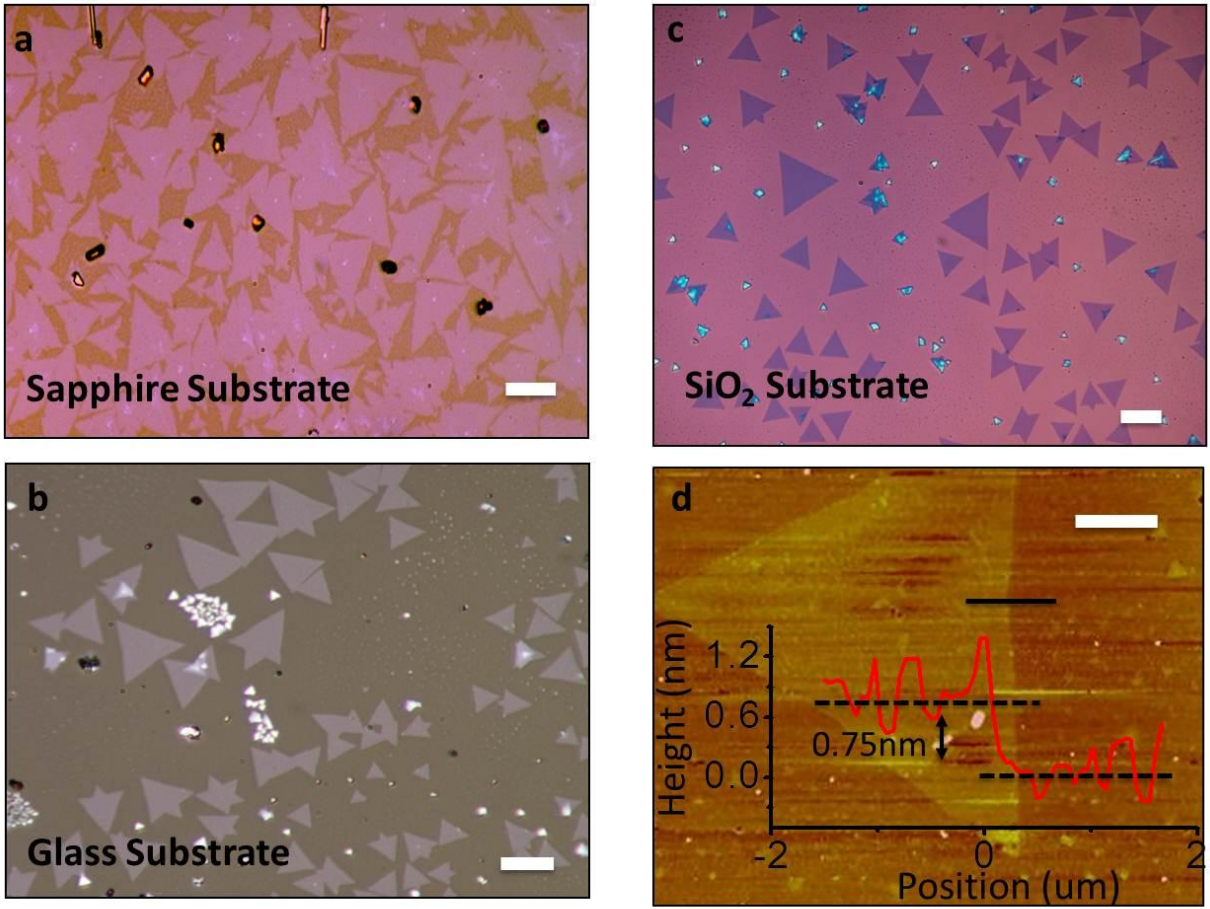

**Figure 3.** Optical microscope images of MoS$_2$ crystallites grown on (a) sapphire, (b) glass, and (c) SiO$_2$/ Si. Scale bar is 10 μm. A typical triangular is further characterized by (d) AFM image with 3 μm scale bar. The inset plot is the height profile along the black line shown image, demonstrating its monolayer thickness.

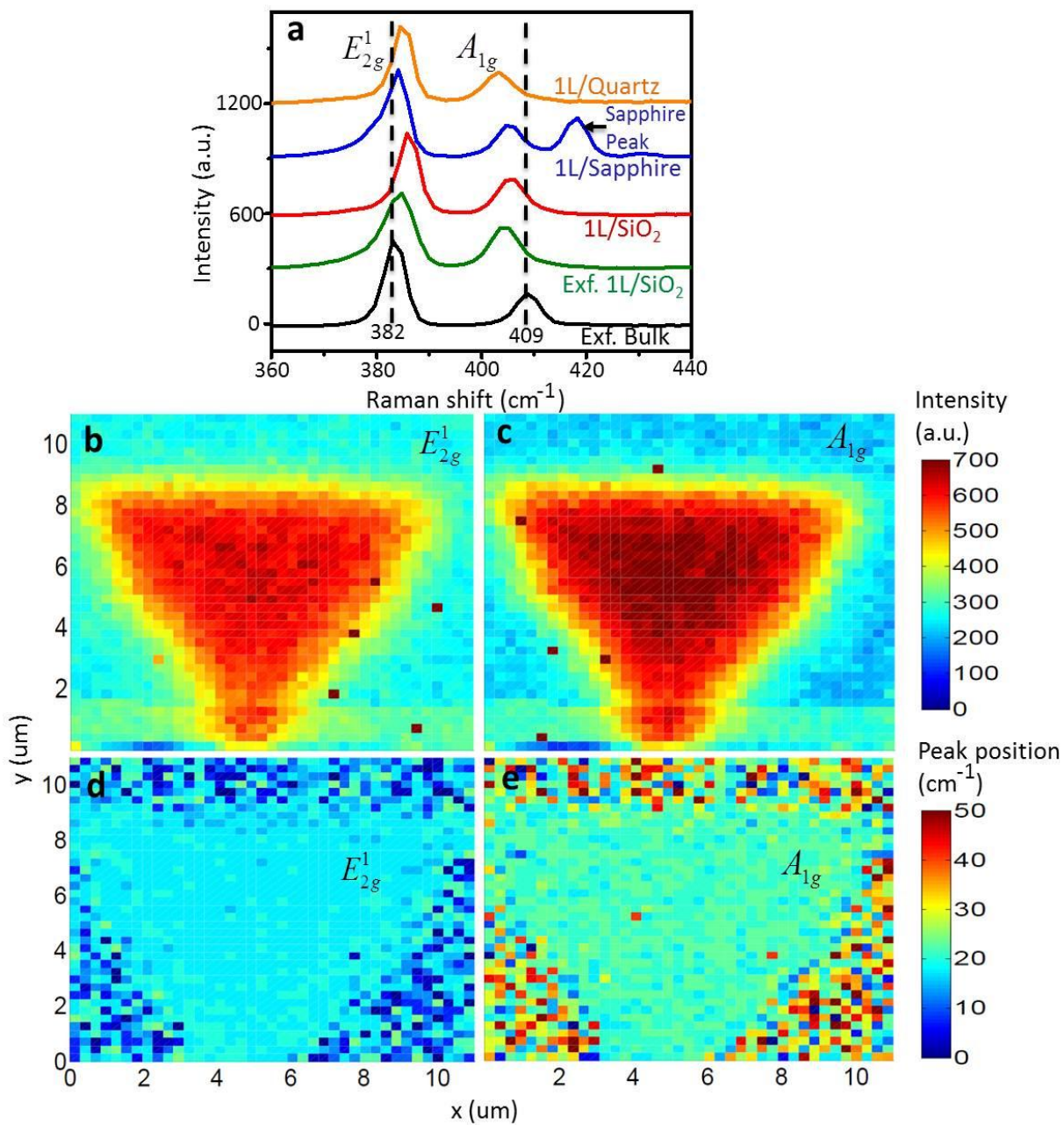

Figure 4. (a) Raman spectra of monolayer MoS$_2$ grown on different substrates, excited by a 514.5 nm laser. For comparison, the spectra from a mechanically exfoliated monolayer and a bulk MoS$_2$ crystal are also shown. (b) and (c), Intensity maps of the two Raman modes, excited by a 532 nm laser line from a typical crystallite. (d) and (e), Corresponding peak position maps.

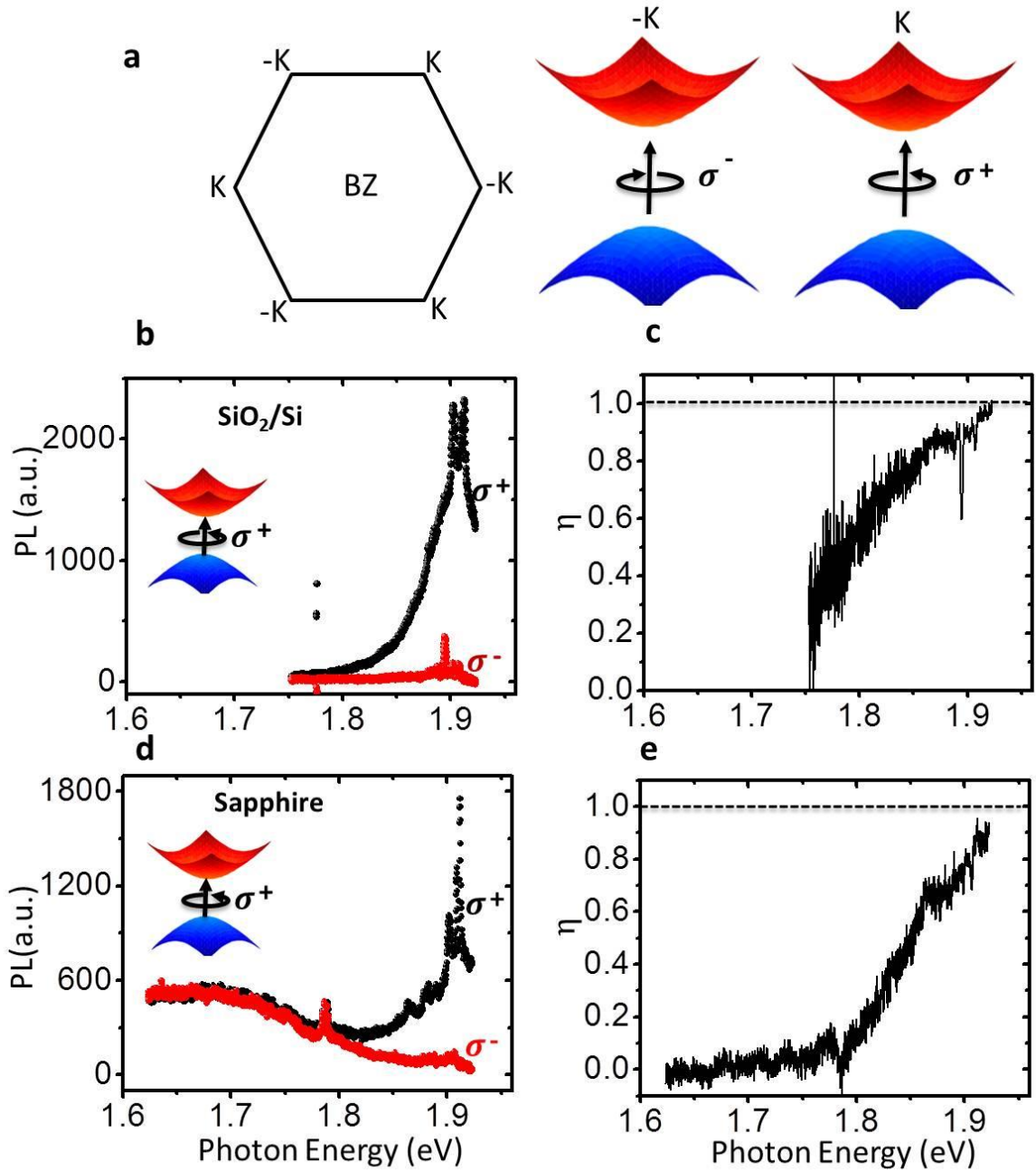

**Figure 5.** (a) Brillouin zone and K-point band edges in monolayer MoS$_2$ indicating the optical selection rule. (b) $\sigma^+$ (black) and $\sigma^-$ (red) components of the PL signal for our monolayer MoS$_2$ on a SiO$_2$/Si substrate, excited by $\sigma^+$ laser light at 632 nm wavelength at 30K. (c) Degree of PL

polarization vs photon energy, calculated from (b). (d) and (e) are corresponding observations on a sample grown on a sapphire substrate. The polarization approaches unity on both substrates.

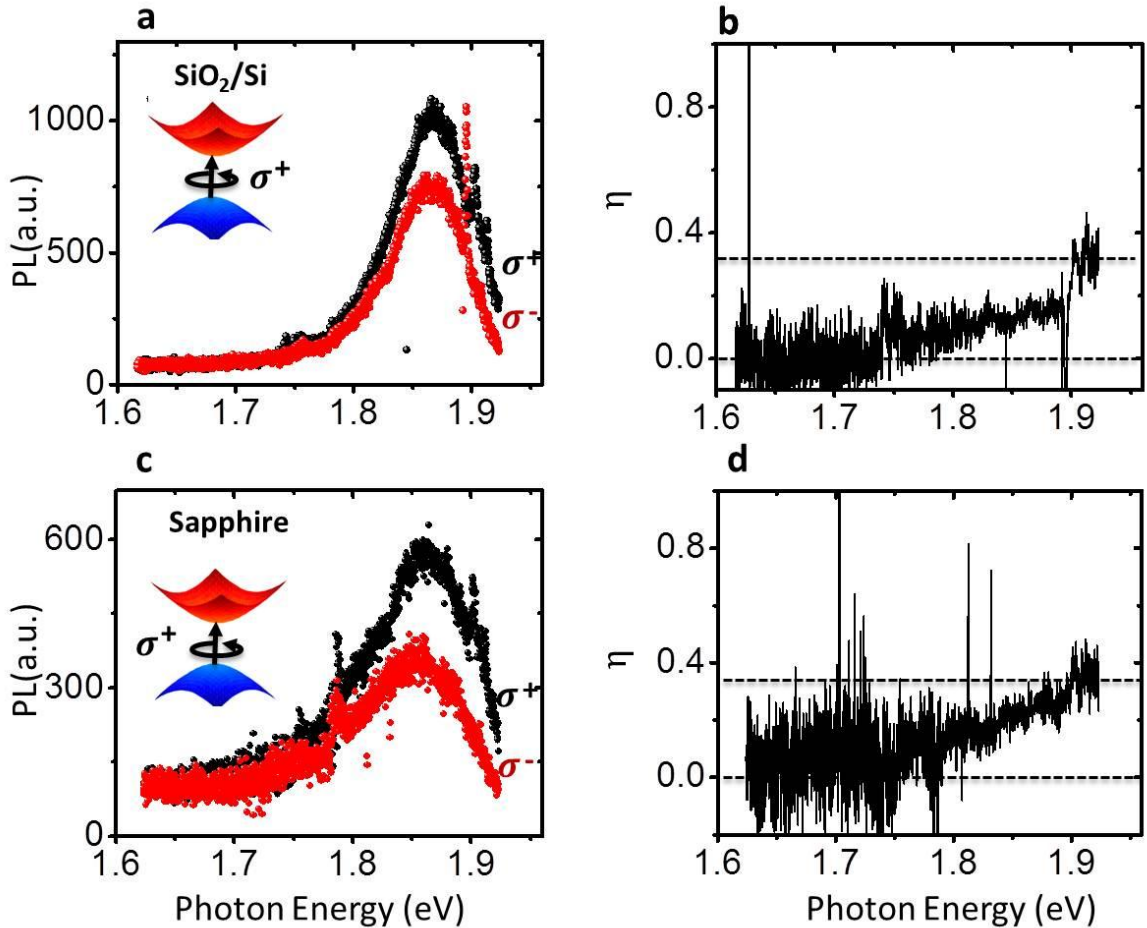

**Figure 6.** (a) $\sigma^+$ and $\sigma^-$ components of the PL for monolayer on $SiO_2$/Si substrate, and (b) degree of PL circular polarization vs photon energy at room temperature. (c) and (d) show similar measurements on a sapphire substrate. Up to ~35% polarization is observed on both substrates.

# Supplementary Material

# Vapor-Solid Growth of High Optical Quality MoS$_2$ Monolayer With Near Unity Valley Polarization

Sanfeng Wu[1*], Chunming Huang[1*], Grant Aivazian[1], Jason Ross, David Cobden[1], Xiaodong Xu[1,2†]

[1] Department of Physics, University of Washington, Seattle, Washington 98195, USA

[2] Department of Material Science and Engineering, University of Washington, Seattle, Washington 98195, USA

\* These authors contribute equally to this work.

† Email address: xuxd@uw.edu

**S1. Growth temperature profile**

**S2. AFM characterization of monolayer grown on sapphire**

**S3. PL polarization under σ⁻ light excitation**

## S1. Growth temperature profile

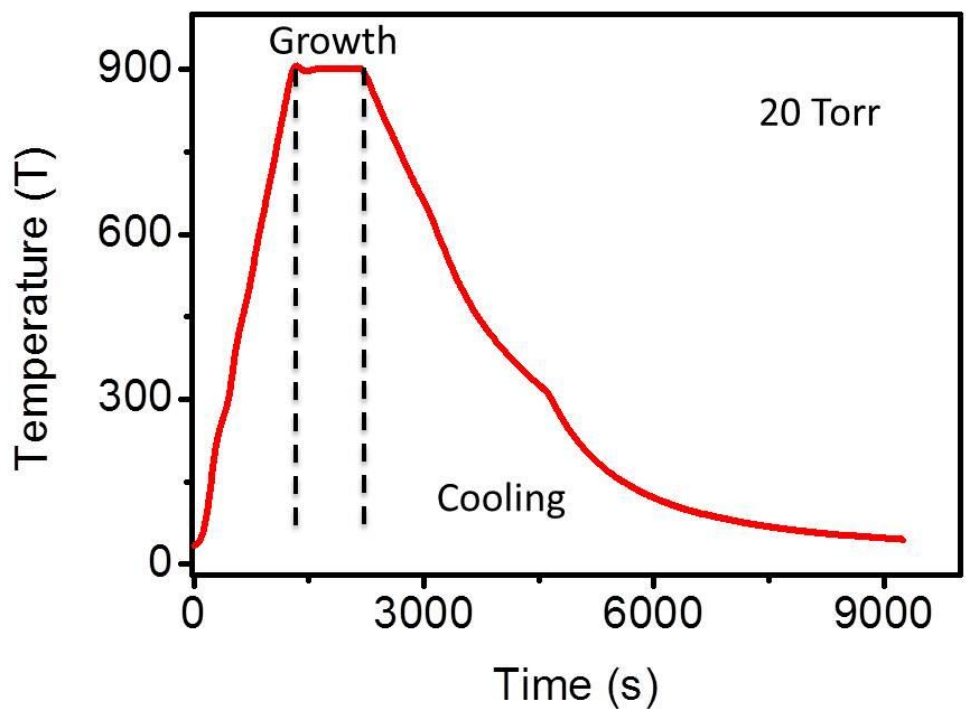

**Figure S1. Temperature at the center of the furnace as a function of time. The temperature is held at ~900 °C for the growth time of ~15 min at 20 Torr of argon.**

## S2. AFM characterization of monolayer MoS$_2$ grown on sapphire

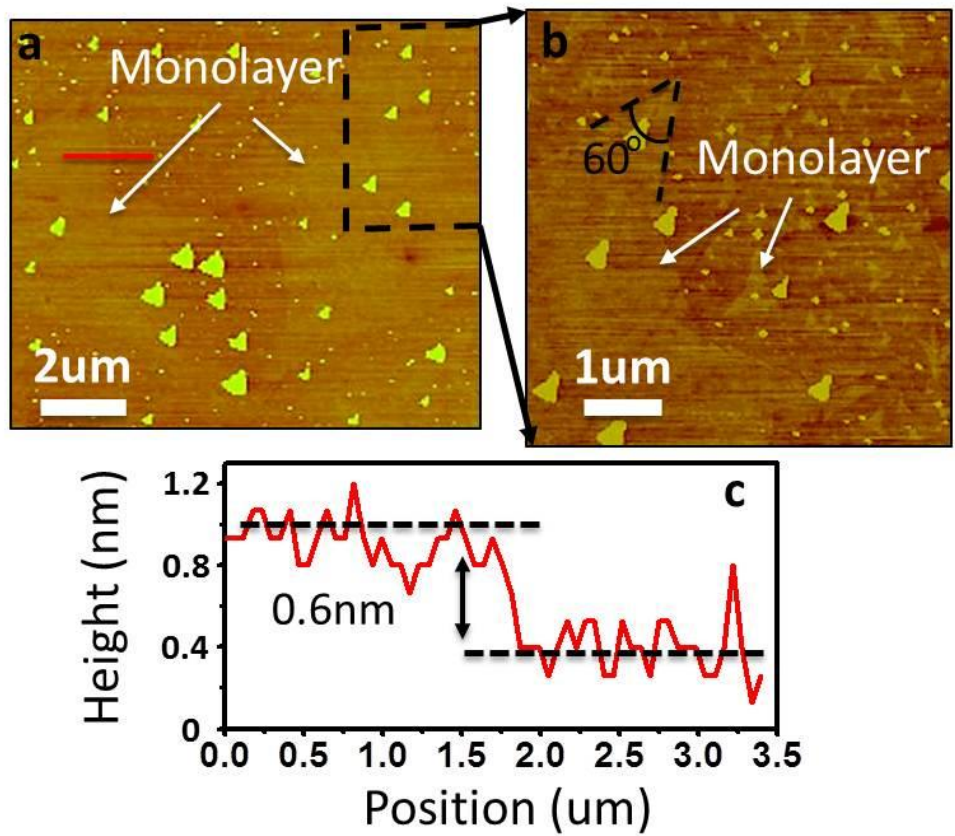

**Figure S2. AFM characterization of monolayer on sapphire substrate.** (a) AFM images of typical as-grown single layer crystallites. (b) Zoom-in image of the area shown in (a). (c) Height profile along the red line in (a), which shows a 0.6 nm height step consistent with a monolayer.

## S3. PL polarization under σ⁻ light excitation

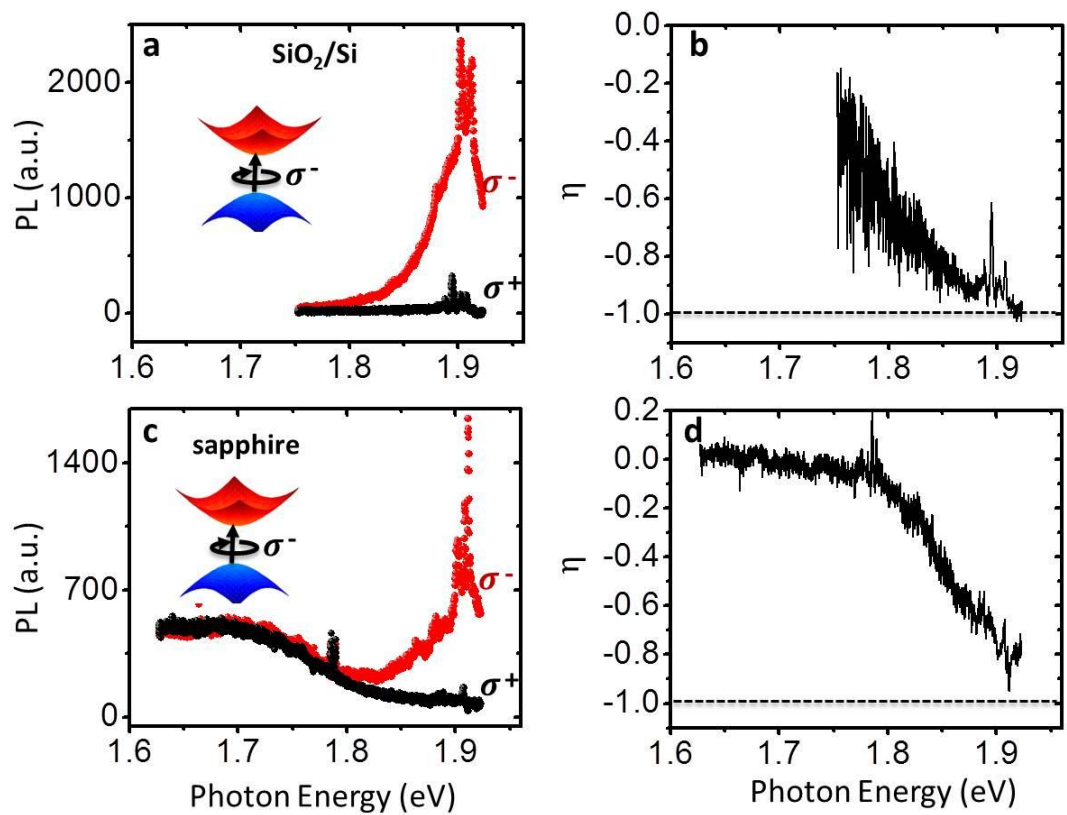

**Figure S3.** (a) $\sigma^+$ (black) and $\sigma^-$ (red) components of the PL spectrum of a crystallite on a SiO$_2$/Si substrate, excited by $\sigma^-$ laser light at 30K, complementary to Fig. 5 in the main text. (b) Degree of PL polarization vs photon energies, calculated from (a). (c) and (d) are corresponding observations on a crystallite grown on a sapphire substrate.